\documentclass[aps,prl,twocolumn,superscriptaddress,groupedaddress,floatfix]{revtex4} 

\usepackage{graphicx} 
\usepackage{caption}
\usepackage{array} 
\usepackage{verbatim} 
\usepackage{amsmath,amsfonts,amssymb,amscd}
\usepackage{amsthm}
\usepackage{thmtools,thm-restate}
\usepackage{enumitem}
\usepackage{algorithm}

\usepackage{hyperref}
\hypersetup{
    bookmarksnumbered=true, 
    unicode=false, 
    pdfstartview={FitH}, 
    pdftitle={}, 
    pdfauthor={}, 
    pdfsubject={}, 
    pdfcreator={}, 
    pdfproducer={}, 
    pdfkeywords={}, 
    pdfnewwindow=true, 
    colorlinks=true, 
    linkcolor=blue, 
    citecolor=blue, 
    filecolor=blue, 
    urlcolor=blue 
}

\newcommand{\eq}[1]{\hyperref[eq:#1]{Eq. (\ref*{eq:#1})}}

\renewcommand{\sec}[1]{\hyperref[sec:#1]{Section~\ref*{sec:#1}}}
\newcommand{\thrm}[1]{\hyperref[thm:#1]{Theorem~\ref*{thm:#1}}}
\newcommand{\lemm}[1]{\hyperref[lemm:#1]{Lemma~\ref*{lemm:#1}}}
\newcommand{\prop}[1]{\hyperref[prop:#1]{Proposition~\ref*{prop:#1}}}
\newcommand{\corr}[1]{\hyperref[corr:#1]{Corollary~\ref*{corr:#1}}}
\newcommand{\fig}[1]{\hyperref[fig:#1]{Figure~\ref*{fig:#1}}}

\newcommand{\ket}[1]{|#1\rangle}
\newcommand{\bra}[1]{\langle#1|}
\newcommand{\bk}[1]{|#1\rangle\langle#1|}

\newcommand{\tth}[0]{\textsuperscript{th}}

\usepackage{xcolor}

\DeclareMathAlphabet{\matheu}{U}{eus}{m}{n}

\DeclareMathOperator{\tr}{tr}

\newcommand{\sop}[1]{{\mathcal #1}}
\newcommand{\PL}[1]{{#1}^{P\!L}}

\newcommand{\ketbra}[2]{|{#1}\rangle\!\langle{#2}|}
\newcommand{\kket}[1]{|{#1}\rangle\!\rangle}
\newcommand{\bbra}[1]{\langle\!\langle{#1}|}

\usepackage{algorithm}
\usepackage{algpseudocode}
\algnewcommand\algorithmicinput{\textbf{Input:}}
\algnewcommand\INPUT{\item[\algorithmicinput]}
\algnewcommand\algorithmicoutput{\textbf{Output:}}
\algnewcommand\OUTPUT{\item[\algorithmicoutput]}
\algnewcommand\And{\textbf{and }}
\algnewcommand\Or{\textbf{or }}
\algrenewcommand\Return{\State \algorithmicreturn{} }

\begin{document}
\title{Experimental demonstration of cheap and accurate phase estimation}
\author{Kenneth Rudinger}
\affiliation{Center for Computing Research, Sandia National Laboratories, Albuquerque, NM 87185, \texttt{kmrudin@sandia.gov}}
\author{Shelby Kimmel}
\affiliation{Joint Center for Quantum Information and Computer Science (QuICS),
University of Maryland, \texttt{shelbyk@umd.edu}}
\author{Daniel Lobser}
\affiliation{Sandia National Laboratories, Albuquerque, NM 87185}
\author{Peter Maunz}
\affiliation{Sandia National Laboratories, Albuquerque, NM 87185}
\begin{abstract}
We demonstrate experimental implementation of robust phase estimation (RPE) to learn the phases of X and Y rotations on a trapped $\textrm{Yb}^+$ ion qubit.  We estimate these phases with uncertainties less than $4\cdot10^{-4}$ radians using as few as 176 total experimental samples per phase, and our estimates exhibit Heisenberg scaling. Unlike standard phase estimation protocols, RPE neither assumes perfect state preparation and measurement, nor requires access to ancillae.  We cross-validate the results of RPE with the more resource-intensive protocol of gate set tomography.
\end{abstract}

\maketitle

\section{Introduction}
As quantum computers grow in size, efficient and accurate methods for
calibrating quantum operations are increasingly important
\cite{EW14,FM15,BWC11,HZT16}. Calibration involves estimating the values of
experimentally tunable parameters of a quantum operation and, if
incorrect, altering the controls to fix the error.

When these tunable parameters are incorrectly set, it causes the system to
experience coherent errors. Coherent errors (versus incoherent errors) are more challenging for error correcting codes to correct
 \cite{SWS16,GSL+16}, making it harder to reach fault-tolerant
thresholds \cite{Kit97,AB97,AB08}. Hence it is important to correct these types
of errors in order to build a scalable quantum computer.  While recent techniques
using randomized compiling \cite{WE15} mitigate the
effects of coherent errors, removing as much of the coherent
errors as possible still gives the best error rates.

Calibration can be challenging to perform without
accurate state preparation and measurement (SPAM) estimates \cite{stark2014,stark2015}. Thus
proper calibration of quantum operations will require \emph{robust} protocols,
that is, protocols that can accurately characterize gate parameters without
highly accurate initial knowledge of SPAM. 

A new technique for calibrating the phases of gate operations is robust phase estimation (RPE) \cite{kimmel15}.  
RPE can be used to estimate the rotation axes and angles of single-qubit
unitaries. Moreover, it is easy to implement (the sequences required are
essentially Rabi/Ramsey experiments), simple and fast to analyze, and can obtain accurate estimates with surprisingly small amounts of data.

RPE has advantages over standard robust characterization procedures when
it comes to the task of calibration. RPE can estimate specific parameters of coherent errors, whereas randomized benchmarking, while robust, can only estimate the magnitude of errors  \cite{KLR+08,MGE12,MGE11,WGHF15,SBM+16}. While compressed sensing approaches can withstand SPAM errors \cite{SKM+11,MCC13}, they do not have the Heisenberg scaling RPE achieves.  There is a simple analytic bound on the size of SPAM errors that RPE can tolerate (namely less than $1/\sqrt{8}$ in trace distance), unlike the robust Bayesian approach of Wiebe et al.,
whose error tolerance is less well-understood. 
\cite{WGFC14}. Lastly, RPE is extremely efficient compared to robust protocols that provide complete reconstructions of error maps, like randomized benchmarking tomography \cite{kimmel14} and gate set tomography (GST) \cite{GST2016}.

Like many other
phase estimation procedures, RPE achieves Heisenberg scaling \cite{kimmel15},
but unlike many other protocols, requires no entanglement such as squeezed states or NOON states \cite{Caves81,K95,KBD04,GLM4,GLM04,MRK+01,LKD02}, requires no ancillae \cite{K95,BFC+07,KLS+14}, and is non-adaptive 
\cite{W95,SCC+11,Higgins2007,FGC13}.

Finally, compared to many tomography and parameter estimation protocols, the
post-experiment analysis of RPE is strikingly simple.  There are no Bayesian updates
\cite{WG15,FGC13,WGFC14}, no optimizations \cite{GST2016,SHF14,SKM+11}, and no
fits to decaying exponentials \cite{MGE11,kimmel14}. Instead, post-processing
involves a dozen lines of pseudo-code, with the most complex operation being an
arctangent (see Supplemental Material for more details).

Here, we provide the first published experimental demonstration of RPE and investigate its performance.  
We use RPE to experimentally extract the phases (rotation angles) of 
single-qubit unitaries. Because we don't know the true values of the parameters, we
benchmark these estimates by comparing to GST, which gives robust,
accurate, and reliable estimates, but which requires much more data
\cite{GST2016}.

We see experimental evidence of Heisenberg scaling in RPE, and we attain an
accuracy of $3.9\cdot10^{-4}$ radians in our phase estimate using only 176 total
samples.  We compare these costs to GST and find that
RPE requires orders of magnitude fewer total gates
and samples to achieve similar accuracies. However, in regimes where
experiments involving long sequences are not accessible, we find GST potentially has better
performance than RPE. Nonetheless, due to its minimal data requirements, ease
of implementation and analysis, and robust estimates of coherent errors, RPE is
a powerful tool for efficient calibration of quantum operations.

\section{Preliminaries}
We consider estimating the parameters $\alpha$ and  $\epsilon$ from the single-qubit gate set \cite{kimmel15}:
\begin{align*}\label{eq:gateset}
\hat{X}_{\pi/2+\alpha}=&\exp\left[-i\left(\left(\pi/2+\alpha\right)/2\right)\hat{\sigma}_X\right],\nonumber\\
\hat{Y}_{\pi/2+\epsilon}(\theta)=&\exp\left[-i\left(\left(\pi/2+\epsilon\right)/2\right)
\left(\cos\theta\hat{\sigma}_Y+\sin\theta\hat{\sigma}_X\right)\right],
\end{align*}
where $\hat{\sigma}_X$ and $\hat{\sigma}_Y$ are Pauli operators, $\alpha$ and $\epsilon$ are rotation errors in the X and Y gates, respectively, and $\theta$ is the size of the off-axis component of the (ideally) Y
gate rotation axis.  There is no off-axis component to the X gates, as we
choose the X axis of the Bloch sphere to be the rotation axis of the X
gate. $\alpha$ and $\epsilon$ are parameters that experimentalists can
typically control with ease.

 In reality, the implemented gates will not be unitary, but instead will be completely positive trace preserving (CPTP) maps.
Nonetheless, these CPTP maps will have rotation angles analogous
to the angles $\alpha$ and $\epsilon$, and in the Supplemental Material, we show
 RPE can extract such angles. For the rest of the paper, with slight abuse of notation, we will use $\alpha$ and $\epsilon$ to refer to these more general CPTP map rotations. 

\begin{figure}[h!]
\includegraphics[width=.4\textwidth]{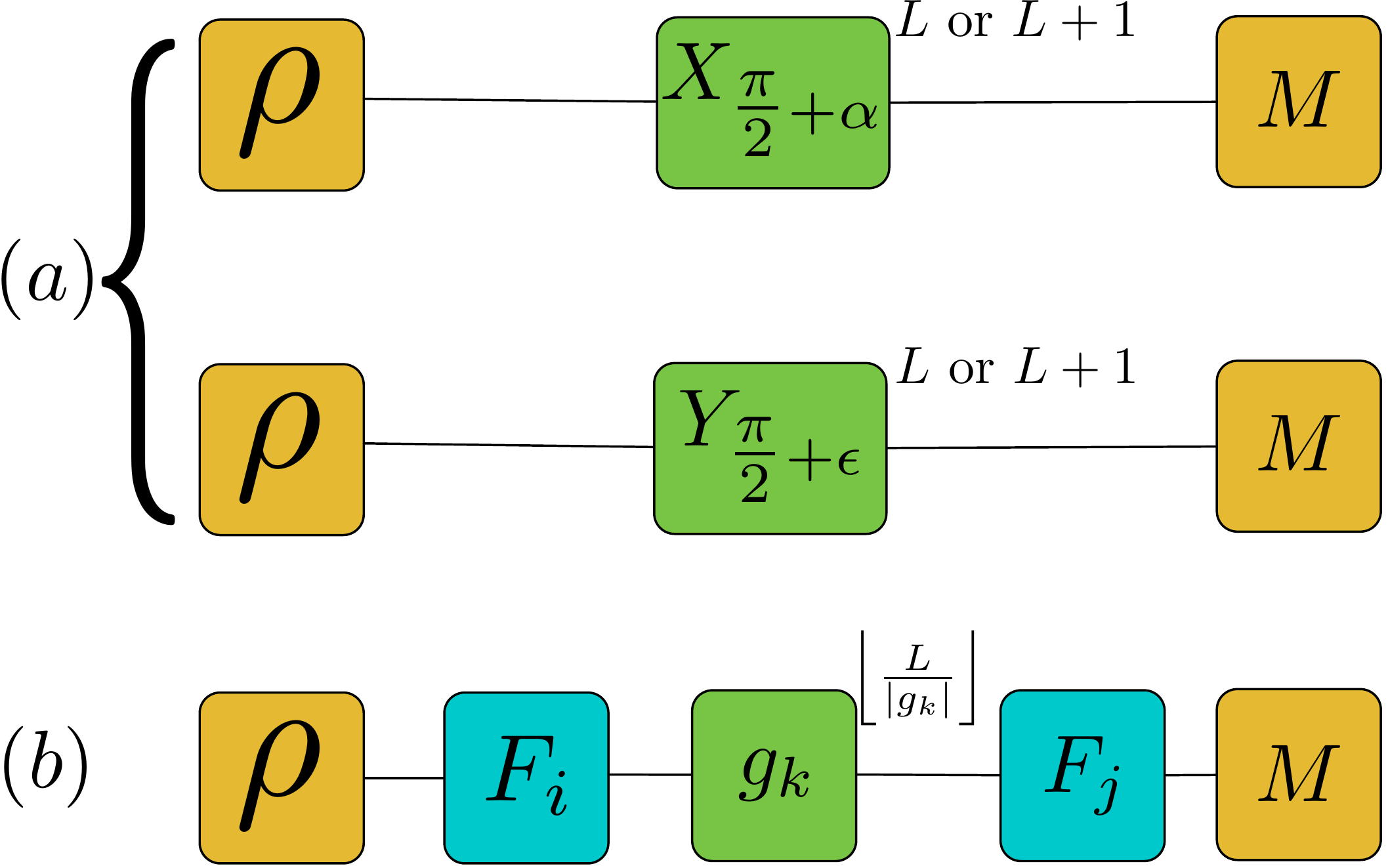}
\caption{\label{fig:circuit}
(Color online)
(a) RPE and (b) GST experimental sequences. Each sequence
starts with the state $\rho$ and ends with the two-outcome 
measurement $M$.  (a) An RPE sequence 
consists of repeating  the gate in question either $L$ or $L+1$ times.  
(b) In GST,
a gate sequence $F_i$ is applied
to simulate a state preparation potentially different from $\rho$.  This is followed by $\lfloor L/|g_k|\rfloor$ applications of a \emph{germ}---a short gate sequence $g_k$ of length $|g_k|$.
Finally, a sequence $F_j$ is applied to simulate a measurement potentially different from $M$.
}
\end{figure}

We use both RPE and GST to extract $\alpha$ and $\epsilon$.  
Fig.~\ref{fig:circuit} gives a schematic description of GST and RPE
circuits. RPE circuits are
essentially Rabi/Ramsey sequences; they consist of state preparation
$\rho$, which for extracting $\alpha$ and $\epsilon$ is assumed to be not too
far in trace distance from $\ketbra{0}{0}$, followed by repeated applications
of the $X$ or $Y$ gate, followed by a measurement operator $M$, which is
assumed to be close in trace distance to $\ketbra{1}{1}$.  (Performing additional,
more complex ``Rabi/Ramsey-like'' sequences allows for RPE to extract $\theta$
as well \cite{kimmel15}; we do not do so here.)

RPE assumes all gates and SPAM are relatively close to ideal, but tolerates
errors.  We use ``additive error'' to denote the maximum bias in the outcome probability
of any single RPE experimental sequence.  This bias can be due to SPAM errors and incoherent errors in
the gates. Additive error can be tolerated as long as it is less than
$1/\sqrt{8}$.

For GST, each sequence consists of a state preparation $\rho$, followed by a
gate sequence $F_i$ to simulate an alternate state preparation. Next a gate
sequence $g_k$ is applied repeatedly. Finally, the measurement $M$ is preceded
by a gate sequence $F_j$ to
simulate an alternative measurement.  We refer to $F_i$ and $F_j$ as state and
measurement fiducials, respectively, and $g_k$ as a germ.  (For more details,
see the Supplemental Material.)

For both RPE and GST, running increasingly longer sequences produces
increasingly accurate estimates. We use $L$ to parameterize the length of the
sequence, as in Fig.~\ref{fig:circuit}. We run sequences with $L \in\{1, 2,
4,8,\dots, L_\textrm{max}\}$, where $L_\textrm{max}$ is chosen based on the
desired accuracy. In RPE, we repeat the gate of interest
either $L$ or $L+1$ times. In GST, we
implement all possible combinations of state fiducials, measurement fiducials,
and germs, with the germ repeated $\lfloor L/|g_k|
\rfloor$ times, where $|g_k|$ is the number of gates in $g_k$ and $\lfloor\cdot\rfloor$ 
denotes the floor function.

We let $N$ be the repetitions (samples taken) of each sequence. We set
$N$ to be the same for all sequences in a single RPE or GST experimental run.
Although this results in slightly non-ideal scaling in the accuracy of our estimate
\cite{HBB+7}, this is a realistic scenario for experimental implementation. 

\begin{figure}[h!]
\includegraphics[width=.49\textwidth]{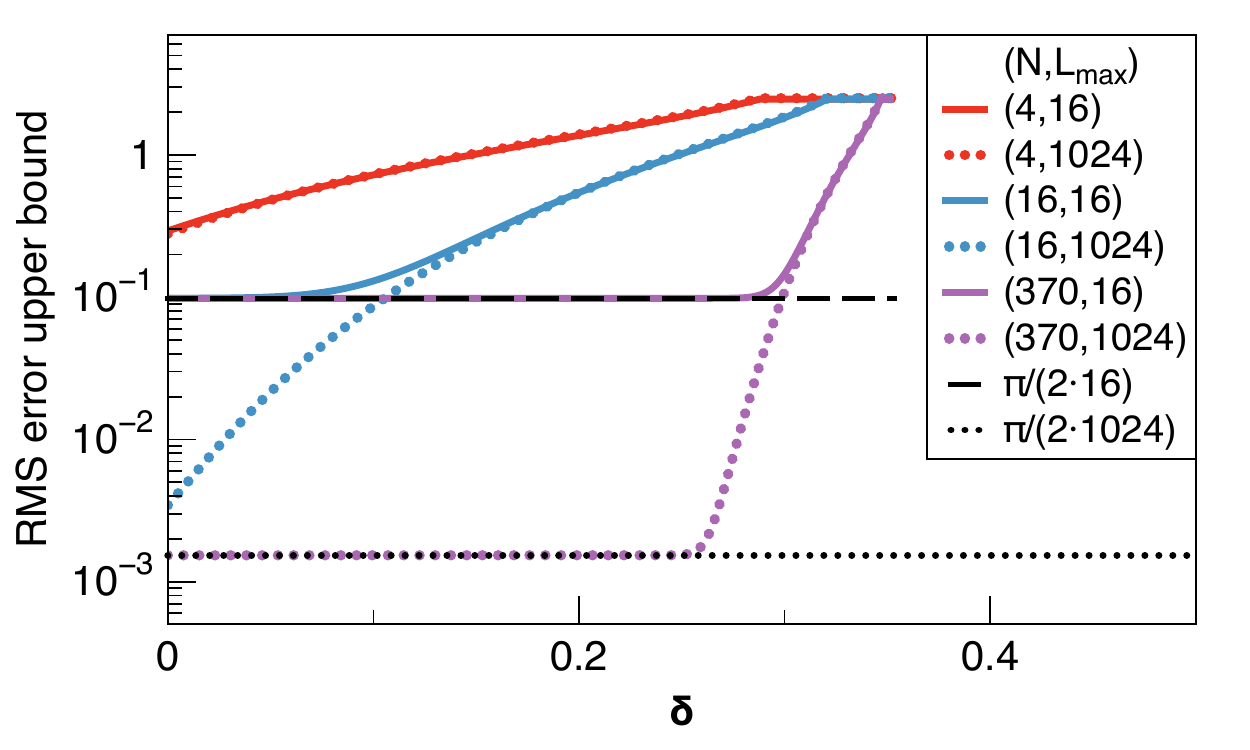}
\caption{\label{fig:Ptot}
(Color online)
Analytic upper bounds on the RMSE of the RPE phase estimate.
Because RPE is potentially biased, the RMSE does not go to zero in the limit of
infinite $N$, but instead, approaches a floor of
$\pi/(2L_{\textrm{max}}).$ Larger additive error $\delta$ produces a larger bias, and thus require larger $N$ and larger $L_{\max}$ to achieve a small RMSE. For example, $N=16$ is not large enough to reach the floor for $L_{\max}=1024,$ but increasing $N$ to $370$ we easily saturate the bound for most values of $\delta$.}
\end{figure}

RPE successively restricts the possible range of the estimated phase
using data from sequences with larger and larger $L$. Inaccuracies result when the procedure restricts to the wrong
range. For larger values of $L_{\max}$, there are more rounds of restricting
the range, and thus more opportunities for failure. By increasing $N$ when
$L_{\max}$ increases, we can limit this probability of failure. Likewise,
a large additive error makes it easier to incorrectly restrict the range, but
again, taking larger $N$ can increase the probability of success. The
interaction between accuracy, $N$, $L_{\max}$, and additive errors is shown in
Fig.~\ref{fig:Ptot}. This graph was created by adapting the analysis of
\cite{kimmel15} to the case of fixed $N$ over the course of an RPE experimental run  
\footnote{In particular, we use Equations 5.7, 5.8, 5.9, and 5.16 from \cite{kimmel15}.}.  
Fig.~\ref{fig:Ptot} shows that, given an additive error $\delta$, there exist good choices for $N$
and $L_{\max}$, provided that $\delta<1/\sqrt{8}$.

A protocol has Heisenberg scaling when the root mean squared error (RMSE) of
its estimate of a gate parameter scales inversely with the number of
applications of a gate. RPE provably has Heisenberg scaling \cite{kimmel15}, and GST
numerically exhibits Heisenberg-like scaling \cite{GST2016}.  In this paper, we
empirically look for scaling in accuracy and precision that scales as
$1/L_{\textrm{max}}$. This is a good proxy (up to log factors) for
Heisenberg scaling.

In practice, experimentalists care less about Heisenberg scaling, and more
about the resources required to achieve a desired accuracy in their estimate.
Therefore we are additionally interested in how large $N$ and $L_\textrm{max}$
should be to attain a desired precision. Assuming time is the key resource, if
experimental reset time is long compared to gate time, $N$ becomes the dominant cost factor. 
On the other hand, if gate time is long compared to experimental reset time, $L_\textrm{max}$ is the dominant factor. 

\section{Experimental results}
Here we will give estimates of $\alpha$. Results for $\epsilon$ are similar and can be found in the Supplemental Material.

We implement GST and RPE on a
single $^{171}$Yb$^+$ ion in a linear surface ion
trap.  The qubit levels are the hyperfine clock states of
the $^2S_{1/2}$ ground state: $\ket{0}=\ket{F=0,m_F
=0},\ket{1}=\ket{F=1,m_F =0}$.  We initialize
the qubit close to the $\ket{0}$ state via Doppler cooling and optical pumping; we 
measure in the computational basis (approximately) via fluorescence
state detection \cite{Olmschenk2007}.  The desired operations are
$X_{\pi/2}$ and $Y_{\pi/2}$. See \cite{GST2016} for experimental details.  
For the numerical analysis in this work, we have used the open-source GST software pyGSTi,
and have extended its capabilities to include RPE functionality \cite{pygsti}.

We take 370 samples of each GST and RPE sequence.
(For details, see Gate Sequences
in Supplemental Material.) We use $L\in\{1,2,4,\dots,1024\}$.
The GST dataset comprises 2,347 unique sequences and
868,390 total samples, while the RPE dataset comprises 44 sequences and 16,280 samples. The RPE dataset further 
disaggregate into disjoint sets of 22 unique sequences and 8,140 samples per phase.

Looking at Fig.~\ref{fig:Ptot}, we see that $N=370$ is larger than necessary for RPE with $L_{\max}=1024$ 
for additive error less than $\sim 0.25$. To simulate experiments with fewer than 370 samples per sequence, we randomly sample (without replacement) from
the experimental dataset, so that the new, subsampled dataset has $N<370$
samples per sequence.

We use several methods to characterize the experimental accuracy of RPE.
First, we apply the analytic bounds on RMSE of Fig.~\ref{fig:Ptot}. We also
compare our subsampled RPE estimates to the GST estimate. Unlike RPE, 
GST is an unbiased estimator \cite{BKG+13}, so going to large $N$ (at a large cost in
resources) gives standard quantum limit scaling. Using the $N=370$ dataset for
GST, we estimate $\alpha-\pi/2=(6.4\pm 4.9)\cdot10^{-5}$; the error bars denote
a 95\% confidence interval derived using a Hessian-based procedure (see
\cite{GST2016} for details). On the other hand, using all RPE data we estimate
$\alpha-\pi/2=1.0\cdot10^{-4}$, with an RMSE upper bound of $\pi/(2\cdot
L_{\max})\approx1.5\times10^{-3}$ (where this bound comes from
Fig~\ref{fig:Ptot} with $N=370$, assuming our additive error is less than $0.25$; this assumption is borne out in the next section).

While the RPE estimate is consistent with the GST result, the accuracy is significantly lower, and we thus take $\alpha_0$, the full
data estimate from GST, to be the ``true'' value of $\alpha$ for
the purposes of benchmarking RPE. In particular, throughout this paper, we calculate experimental RMSE by 
comparing the mean estimate from 100 subsampled datasets to $\alpha_0$.

\subsection{Heisenberg Scaling from RPE}

\begin{figure}[h!]
\includegraphics[width=.49\textwidth]{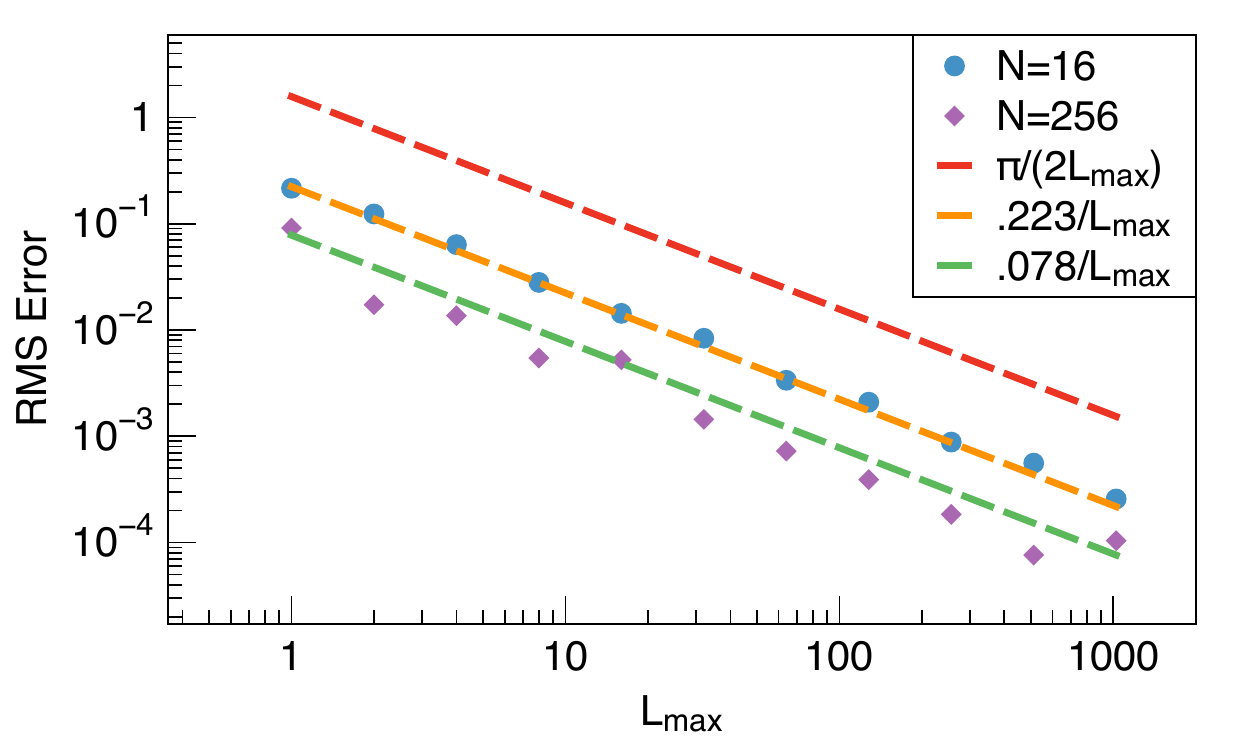}
\caption{\label{fig:2}
(Color online) RSME versus $L_{\max}$ for RPE estimates of $\alpha$ from 100 subsampled datasets of size $N=16$ and $N=256$.   
While analytic bounds are at best $\pi/(2L_{\max})$, we see this can be pessimistic.  When the additive errors, which can bias the RPE estimate, are sufficiently small, increasing $N$
improves RMS accuracy.}
\end{figure}

 To look for Heisenberg scaling in RPE estimates, we perform RPE on 100
 subsampled datasets for $L_\textrm{max}\in\{1,2,4,\dots,1024\}$ with
 $N\in\{16,256\}$. We see Heisenberg-like scaling in the experimental RMSE in
 Fig.~\ref{fig:2}. We also plot $\pi/(2L_{\max})$, which is the analytic upper
 bound if sufficient samples are taken to compensate for additive error. We see
 that in practice, the analytic bounds can be pessimistic. Moreover, we see
 that while the experimental RPE accuracy {\it is} sensitive to $N$, increasing
 $N$ to $256$ from $16$ does not dramatically improve the RMSE, improving the
 scaling to $.078/L_{\max}$ from $.223/L_{\max}$. Instead, as expected, large
 increases in accuracy are obtained by moving to larger $L_{\max}$. This
 Heisenberg-like scaling is especially important for regimes where the time to
 implement the gate sequence is long relative to SPAM time.

We believe our experimentally derived bounds are significantly better than our
analytic bounds in part because our system is well calibrated. The analytic
bounds give a worst-case analysis that accounts for bias caused by adversarial additive error,
but RPE is effectively unbiased for our system, up to the accuracy we achieve.

\subsection{Comparison to GST}

Because RPE can be biased, increasing $N$ cannot improve the RMSE below $\pi/(2L_{\max})$ in the worst case (see Fig.~\ref{fig:Ptot} and \cite{kimmel15}).
However since GST is unbiased, it always
benefits from increasing $N.$

\begin{figure}[h!]
\includegraphics[width=.49\textwidth]{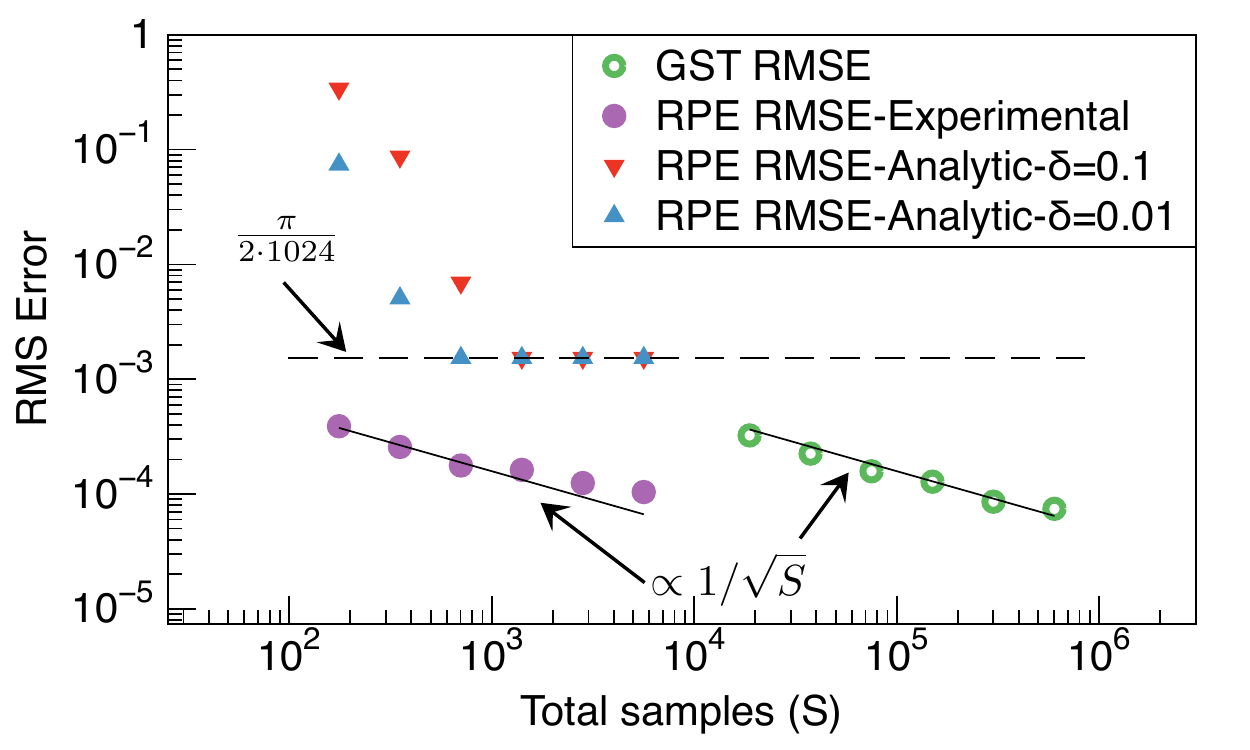}
\caption{\label{fig:tab}
(Color online) Scaling of RMSE of estimates of $\alpha$ as a function of total samples (S), with $L_{\max}=1024$, and $N\in\{8,16,\dots,256\}$. The data point furthest left in each sequence corresponds to $N=8$, and the furthest right to $N=256.$
Analytic bounds are derived using the techniques of Fig \ref{fig:Ptot}.
Experimental data points take the RMSE of 100 subsampled datasets for both RPE and GST.
While the analytic bounds converge to $\pi/2048$,
we see standard quantum limit scaling (i.e., error scaling $\propto1/\sqrt{S}$) of RPE experimental estimates. As discussed in the text, this is because our experimental device has very low additive error, and thus the RPE estimates are essentially unbiased, and can achieve greater accuracy with increasing number of samples. GST estimates also exhibit standard quantum limit scaling.}
\end{figure}

We investigate this effect in Fig.~\ref{fig:tab}. We plot the RMSE for experiments with fixed $L_{\max}=1024$, but
$N\in\{8,16,\dots, 256\}$. Analytic bounds for RPE are
derived using the same method as in Fig~\ref{fig:Ptot}.
Experimental bounds for GST and RPE are derived from comparing the estimates of
100 subsampled datasets to $\alpha_0.$

While the analytic RPE bounds do not improve with increasing $N$, the subsampled RPE and GST datasets show standard quantum limit scaling. We expect this for GST, because GST is unbiased. In the case of RPE our experimental system happens to have very small additive error, and so is only very slightly biased. In this case, we expect to see improving estimates with increasing $N$ until our accuracy is about the same size as our bias. Fig.~\ref{fig:tab} tells us that for systems with relatively large additive error, where large $N$ is feasible but large $L_{\max}$ is not, GST can provide more accurate results. 

\begin{figure}[h!]
\includegraphics[width=.49\textwidth]{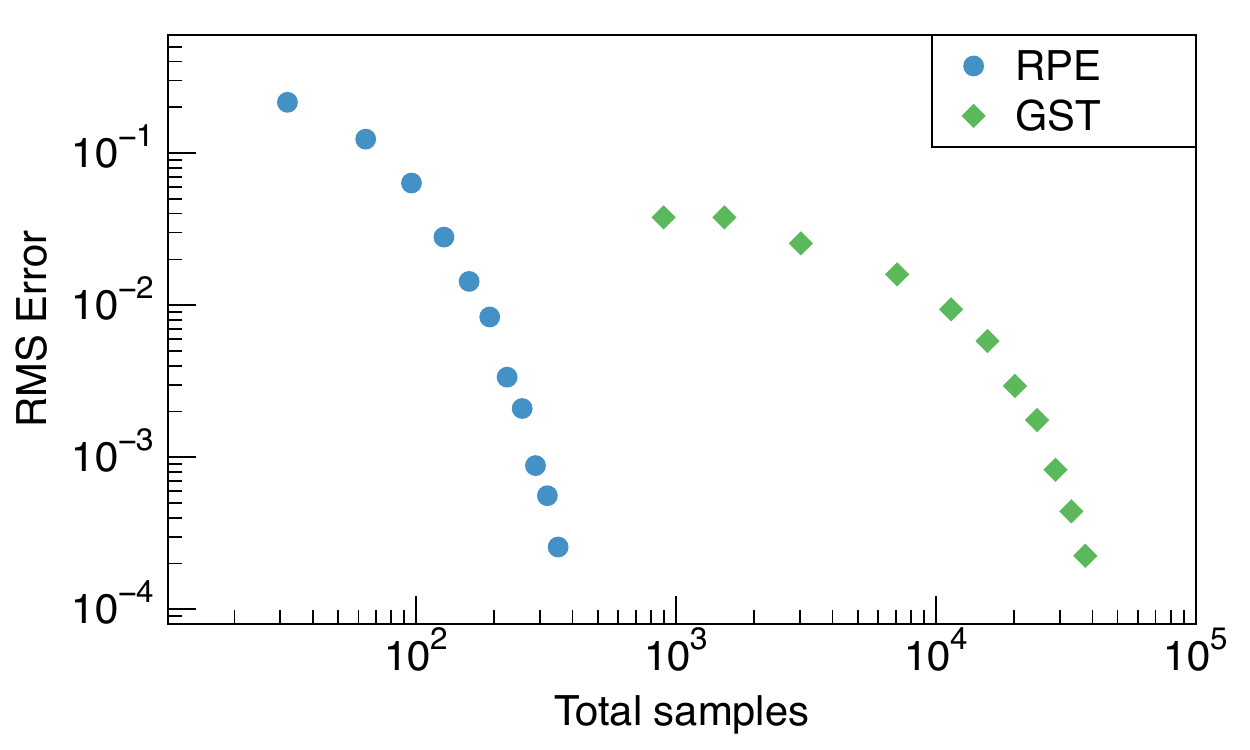}
\caption{\label{fig:3}
(Color online) RMSE for RPE and GST estimates of $\alpha$ versus total number of samples, using 100 subsampled datasets with $N=16$.  Each sequential data point corresponds to setting $L_{\max}\in\{1,2,3,\dots,1024\}$.  RPE achieves the same level of accuracy as GST using far fewer resources.
}
\end{figure}

However, we see in Fig.~\ref{fig:tab} that GST pays a substantial cost
relative to RPE in required number of total samples (i.e., number of samples per sequence $N$ times total number of sequences). 
In Fig.~\ref{fig:3}, we compare
the number of gates and samples which RPE and GST each require to achieve a desired
accuracy, by analyzing 100 subsampled datasets with fixed $N=16$ and varying $L_{\max}\in\{1,2,4,\dots, 1024\}.$
We see that RPE can achieve similar accuracy to GST while using at least an order
of magnitude fewer total samples. 

For our system, acquiring the entire RPE
and GST datasets took 10.8 minutes and 12.1 hours, respectively,
and total experimental time scales linearly with $N$.
Thus we note that had our actual data acquisition rate been $N=16$, it would have taken 28 s to acquire
that RPE dataset and about 31 minutes to acquire the GST dataset.
As for analysis time, a single RPE dataset can be analyzed in
about 0.05 s on a modern laptop.  GST analysis takes about 20 s \footnote{All analyses performed on a 2014 MacBook Pro
2.5 GHz Intel Core i7 machine.}.  All datasets and analysis notebooks are available online \cite{RPESuppData}.


\section{Conclusions}
We show that robust phase estimation successfully
estimates the phases of single-qubit gates, yielding results that are consistent with the
full tomographic reconstruction of gate set tomography and also
exhibits Heisenberg-like scaling in accuracy. In particular, an individual phase may be estimated with a root mean squared error of
$3.9\cdot10^{-4}$ with as few as 176 total samples.

Hence, RPE is a strong choice for diagnosing and calibrating single-qubit operations.
It would be interesting to investigate whether the techniques of RPE can be applied to assessing other errors in
single-qubit gate operations in a fast and accurate manner.

\section{Acknowledgements}
Sandia National Laboratories is a multi-program laboratory
managed and operated by Sandia Corporation,
a wholly owned subsidiary of Lockheed Martin Corporation,
for the U.S. Department of Energy's National Nuclear
Security Administration under contract DE-AC04-
94AL85000.  SK is funded by the Department of Defense.
The authors thank Robin Blume-Kohout and Nathan Wiebe
for helpful conversations, and Erik Nielsen for
extensive software support.  This research was funded, 
in part, by the Office of the Director of National Intelligence (ODNI), 
Intelligence Advanced Research Projects Activity (IARPA). 
All statements of fact, opinion or conclusions contained herein 
are those of the authors and should not be construed as 
representing the official views or policies of IARPA, the ODNI, 
or the U.S. Government.

\section{Supplemental Material} 

\subsection{Robust phase estimation on CPTP maps}
\label{sec:theory}
In RPE, \cite{kimmel15}, the gates to be analyzed  are assumed to be close to
some unitaries. Then RPE allows estimation of the error parameters
 of those unitaries (see Eq.~1). However, there is ambiguity in this
formulation, because given a full description of completely positive and trace
preserving (CPTP) map $\sop E$, there is not a unique unitary  associated to
this map. This might  make it difficult to compare RPE and GST, since GST
produces an estimate for a complete CPTP map. We now show that given a CPTP map
$\sop E$ on a single qubit, RPE can extract the phase of the imaginary eigenvalues
of that map.

We will use the Pauli-Liouville representation of CPTP maps, states and measurements (e.g. \cite{KR01}). \textcolor{black}{
Let $P_i=\sigma_i$ (the single-qubit Pauli matrices) and let $P_0$ be the 2-by-2 identity matrix.}
Then for a single-qubit CPTP map $\sop E$, the Pauli-Liouville representation 
$\PL{\sop E}$ is given by
\begin{align}
\PL{\sop E}=\sum_{i,j=0}^{3}\frac{\tr\left(\sop E(P_i)P_j\right)}{2}\ketbra{i}{j}.
\end{align}

In the Pauli-Liouville representation, a single qubit density matrix $\rho$ is given by the vector
$\kket{\rho}$ where 
\begin{align}
\kket{\rho}=\sum_{i=0}^3\frac{1}{\sqrt{2}}\tr(\rho P_i)\ket{i},
\end{align}
and a positive measurement operator $M$ is given by $\bbra{M}$ where
\begin{align}
\bbra{M}=\sum_{i=0}^3\frac{1}{\sqrt{2}}\tr(M P_i)\bra{i}.
\end{align}

As a consequence of these definitions, we have that $\tr(M\sop E(\rho))=\bbra{M}\PL{\sop E}\kket{\rho}.$ Thus, as in GST, using an invertible $4\times 4$ matrix $S$, we can transform all states $\kket{\rho}$, maps $\PL{\sop E}$, and measurements $\bbra{M}$
as
\begin{align}
\PL{\sop E}\rightarrow& S^{-1}\PL{\sop E}S\nonumber\\
\kket{\rho}\rightarrow& S^{-1}\kket{\rho}\nonumber\\
\bbra{M}\rightarrow& \bbra{M}S,
\end{align}
and not impact any observables.

For single-qubit CPTP maps, $\PL{\sop E}$ is a real $4\times 4$ matrix~\cite{KR01} with two real
eigenvalues (one of which has value 1) and two complex eigenvalues
(which are complex conjugates of each other, by the complex conjugate root theorem) \footnote{It is possible
that all eigenvalues are real if the map is purely
depolarizing/dephasing, but we will ignore this case.}. 
Let  $re^{\pm i\phi}$ be the phases of the complex eigenvalues of a
map $\PL{\sop E}$.
Using a similarity transformation $S_{\sop E}$ (in particular, the matrix
whose columns are the right eigenvectors of $\PL{\sop E}$), we can transform $\PL{\sop E}$ to
${\PL{\sop E}}'=S_{\sop E}^{-1}\PL{\sop E} S_{\sop E}$, where ${\PL{\sop E}}'$ has the form
\begin{align}
{\PL{\sop E}}'=\left(
\begin{array}{cccc}
1&0&0&0\\
0&re^{i\phi} & 0 & 0\\
0&0 & re^{-i\phi} & 0\\
0&0&0&d\\
\end{array}
\right).
\end{align}

Now suppose we can prepare the state $\textcolor{black}{\rho_x\approx\ket{+}\!\bra{+}}$, and
make measurements $M_x$ \textcolor{black}{and} $M_y$ (measurements in the $\sigma_x$ and $\sigma_y$ bases, respectively).  \textcolor{black}{By construction, we assert that, 
under the same similarity 
transformation $S_{\sop E}$,} we have
\begin{align}\label{eq:CPTP-RPE}
 S_{\sop E}^{-1}\kket{\rho_x}=
(1/\sqrt{2},1/2,1/2,0)^T+\kket{\delta_{\rho_x}}\nonumber\\
 \bbra{M_x}S_{\sop E}=
(1/\sqrt{2},1/2,1/2,0)+\bbra{\delta_{M_x}}\nonumber\\
 \bbra{M_y}S_{\sop E}=
(1/\sqrt{2},-i/2,i/2,0)+\bbra{\delta_{M_y}}.
 \end{align}   
 \textcolor{black}{We may assert the above because any errors introduced by $S_{\sop E}$ get absorbed into the $\delta$ terms.  
 (Physically $\kket{\delta_{\rho_x}}$, $\bbra{\delta_{M_x}}$ and $\bbra{\delta_{M_y}}$ correspond to additive errors present in the state preparation and measurement operations.)}
Then we have
\begin{align}
\bbra{M_x}(\PL{\sop E})^k\kket{\rho_x}=\frac{1}{2}(1+\cos(k\phi))+\delta_x^k\nonumber\\
\bbra{M_y}(\PL{\sop E})^k\kket{\rho_x}=\frac{1}{2}(1+\sin(k\phi))+\delta_y^k
\end{align}
where $(\PL{\sop E})^k$ signifies acting with $\sop E$ repeatedly $k$ times, and $\delta_x^k$ and $\delta_y^k$ depend on $r$ as well as $\kket{\delta_{\rho_x}}$, $\bbra{\delta_{M_x}}$, and $\bbra{\delta_{M_y}}$.

Comparing Eq.~\ref{eq:CPTP-RPE} with Eq. (I.1)-(I.2) of \cite{kimmel15}, we see that given these two types
of measurements, RPE can be used to learn $\phi$, assuming $\delta^k_x$ and $\delta^k_y$ are not too large.  
Thus we can directly
compare estimates of rotation angles obtained by GST or by RPE.

\subsection{Gate sequences}

Detailed explanations for the choice of gate sequences used for RPE and GST are given in 
\cite{kimmel15} and \cite{GST2016}, respectively.  Here we simply provide
complete descriptions of the gate sequences used.

Before proceeding further, we describe two notational conventions:
We denote the $X_{\pi/2}$ gate as $G_x$, and $Y_{\pi/2}$ gate as $G_y$.  Additionally,
sequences are listed in operation order, not matrix multiplication order, so the sequence
$G_xG_y$ means ``apply the $X_{\pi/2}$ gate, and then apply the $Y_{\pi/2}$ gate''.

Both RPE and GST rely on gate sequences that have a well-defined structure.
For GST, each sequence is of the following form:
\begin{enumerate}
\item Prepare a fixed input state.
\item Apply a short gate sequence (called a \emph{fiducial preparation}, denoted $F_i$) to simulate a particular state preparation.
\item Apply a short gate sequence (called a \emph{germ}, denoted $g_k$) $\lfloor L/|g_k|\rfloor$ times, where $|g_k|$ is the number of gates in the germ, and $L\in \mathbb{Z}^+$ is the sequence length. 
\item Apply a short gate sequence (called a \emph{fiducial measurement}, denoted $F_j$) to simulate a particular measurement operation.
\item Perform and record the outcome of a fixed measurement.
\end{enumerate}

RPE uses fiducial sequences and germs as well.  However, the fiducial sequences are \emph{not} independent of the germ under consideration, as we will describe in more detail when we discuss the specific RPE gate sequences.

We divide experiments into \emph{generations}, labeled by $m\in\{0,1,\dots\}.$ Sequences in generation $m$ have sequence length $L=2^{m}$. For example, for the $m=3$ generation, the underlying sequence (modulo 
fiducials) for the germ $G_x$ is simply $G_x^8$; for the germ $G_x G_y$, it is $(G_xG_y)^4$, and for the germ 
$G_yG_xG_yG_xG_xG_x$, it is just $G_yG_xG_yG_xG_xG_x$.

In GST, for each generation and each germ, sequences are run with every
possible pairing of fiducial state preparation and measurement. That is, if
there are $f_p$ and $f_m$ unique fiducial preparations and measurements respectively, then
there are $f_p\cdot f_m$ unique sequences for a particular germ for a
particular generation.

In our experiments, there are 11 generations in total (ranging from $m=0$ to
$m=10$).  Additionally, our target preparation operation is always $\bk{0}$,
and our target measurement operation is always $\sigma_z$.

\subsubsection{GST fiducials}

The preparation and measurement fiducials that we use for GST are, conveniently, identical.  They correspond to mapping both the state preparation and measurement vectors to the six antipodal points on the Bloch sphere that intersect with the X, Y, and Z axes.  Therefore, each germ at each generation generates 36 different sequences.  The fiducials are:\\
\begin{enumerate}[nolistsep]
\item \{\}  (The null sequence; do nothing for no time.)
\item $G_x$
\item $G_y$
\item $G_x G_x$
\item $G_x G_x G_x$
\item $G_y G_y G_y$
\end{enumerate}

\subsubsection{GST germs}

The germs we use for GST in this Letter are:\\
\begin{enumerate}[nolistsep]
\item $G_x$
\item $G_y$
\item $G_xG_y$
\item $G_yG_yG_yG_x$
\item $G_yG_xG_yG_xG_xG_x$
\item $G_yG_xG_yG_yG_xG_x$
\item $G_yG_yG_yG_xG_yG_x$
\item $G_xG_xG_yG_xG_yG_y$
\end{enumerate}

\subsubsection{RPE germs and fiducials}

The fiducials and germs used in an RPE sequence will depend on both the quantity being estimated, 
and the native fixed input and fixed measurement. In particular, for our
experimental system, we believe that the fixed input state is close to $\bk{0}$ and the fixed
measurement is close $\sigma_z$. Then for $\alpha$ (the
amount of over- or under-rotation in $G_x$), the germ is $G_x$, state preparation
is always the empty fiducial $\{\}$, and there are two measurement fiducials, the empty fiducial $\{\}$ and
the gate $G_x$; for
$\epsilon$ (the amount of over- or under-rotation in $G_y$), the germ is
 $G_y$, state preparation
is always the empty fiducial $\{\}$, and there are two measurement fiducials, the empty fiducial $\{\}$ and
the gate $G_y$;

Therefore, every RPE sequence we apply has one of the following forms:

\begin{enumerate}[nolistsep]
\item ${G_x}^{2^m}$
\item ${G_x}^{2^m+1}$
\item ${G_y}^{2^m}$
\item ${G_y}^{2^m+1}$
\end{enumerate}
for $m\in\{0,\cdots,10\}$.
\subsection{RPE Algorithm}\label{sec:alg}

We use the following is the algorithm that takes raw data counts from a robust phase estimation experiment, and returns an estimate of the phase.  An open-source implementation of this protocol is available online \cite{pygsti}.

\begin{algorithm}[H]
\caption{}
\label{alg:RPE}
\begin{algorithmic}[1]
    \INPUT 
    \\$\vec{M}\in [\mathbb{Z}^+]^n$, a vector whose $i\tth$ element $M_i$ is the number of repetitions (samples) of the $i\tth$ experiment.
    \\ $\vec{x}\in [\mathbb{Z}^+]^n$, a vector whose $i\tth$ input $x_i$ is sampled from a binomial distribution $B(\cos(2^{i-1}\phi)/2+1/2+\delta_{i,x}, M_i)$, with $\delta_{i,x}\in[0,1]$ for all $i$.
    \\ $\vec{y}\in [\mathbb{Z}^+]^n$, a vector whose $i\tth$ input $y_i$ is sampled from a binomial distribution $B(\sin(2^{i-1}\phi)/2+1/2+\delta_{i,y},M_i)$, with $\delta_{i,y}\in[0,1]$ for all $i$.
    \OUTPUT Estimate $\hat{\phi}\in [-\pi,\pi]$ of $\phi$
    \Function{RobustPhaseEstimation}{$\vec{M}$, $\vec{x}$, $\vec{y}$} 
    \State $\textrm{Estimate}=0$ \Comment{Initial estimate could be any value in $[-\pi,\pi]$; algorithm would be unaffected.}
    \For{$i=1$ to $n$}
    	\State $L=2^{i-1}$
    	\State $\textrm{CurrentPhase}=\arctan2((x_i-M_i/2)/M_i,-(y_i-M_i/2)/M_i))/L$ \Comment{Calculate the remainder of the estimate mod $1/L$.}
    	\While{$\textrm{CurrentPhase}<(\textrm{Estimate}-\pi/L)$}
    		\State $\textrm{CurrentPhase}=\textrm{CurrentPhase}+2\pi/L$  \Comment{If smaller than allowed principle range, increase until in range.}
    	\EndWhile
    	\While{$\textrm{CurrentPhase}>(\textrm{Estimate}+\pi/L)$}
    		\State $\textrm{CurrentPhase}=\textrm{CurrentPhase}-2\pi/L$  \Comment{If larger than allowed principle range, decrease until in range.}
    	\EndWhile
    	\State $\textrm{Estimate}=\textrm{CurrentPhase}$
    \EndFor
    \Return $\textrm{Estimate}$
    \EndFunction
\end{algorithmic}
\end{algorithm}

\subsection{Results for $\epsilon$}
We now present our experimental results for the rotation angle $\epsilon$, corresponding
to Figs. 3 and 5.  We see that the $\epsilon$ estimate and error bar behaviors
are both qualitatively and quantitatively similar to the $\alpha$ behavior.  In particular, we observe
$1/L_{\max}$ scaling in the RPE estimates for $\epsilon$ at $N$ as low as 16, and the observed
RMSE scaling constant is below that guaranteed by RPE theory.  Additionally, we find that, using 
the full $N=370$ dataset, GST returns the estimate $\epsilon-\pi/2 = 2.7\cdot10^{-5}\pm3.5\cdot10^{-5}$.  
RPE provides a consistent estimate of $\epsilon-\pi/2 = 9.9\cdot10^{-5}$, with an RMSE upper bound of 
$\pi/(2\cdot L_{\max})\approx1.5\times10^{-3}$.

\begin{figure}[htbp!]
\includegraphics[width=0.49\textwidth]{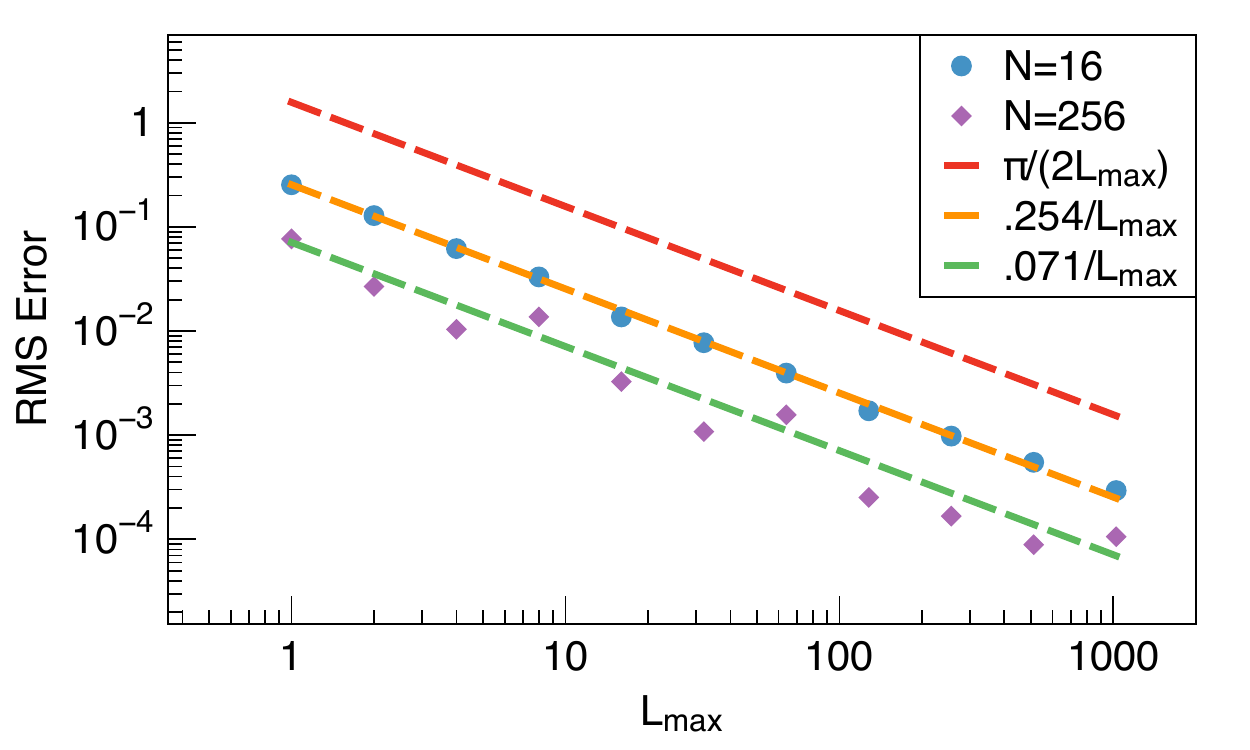}
\caption{\label{fig:Epsilon1}
(Color online) Root mean squared error vs $L_{\max}$ for RPE estimates of the angle $\epsilon$ from subsampled datasets
with $N=16$ and $N=256$.  100 subsampled datasets are used for each value of $N$.  This plot is
analogous to the plot shown in Fig. 3; for further details, see caption of that figure.}
\end{figure}

\begin{figure}[htbp!]
\includegraphics[width=0.49\textwidth]{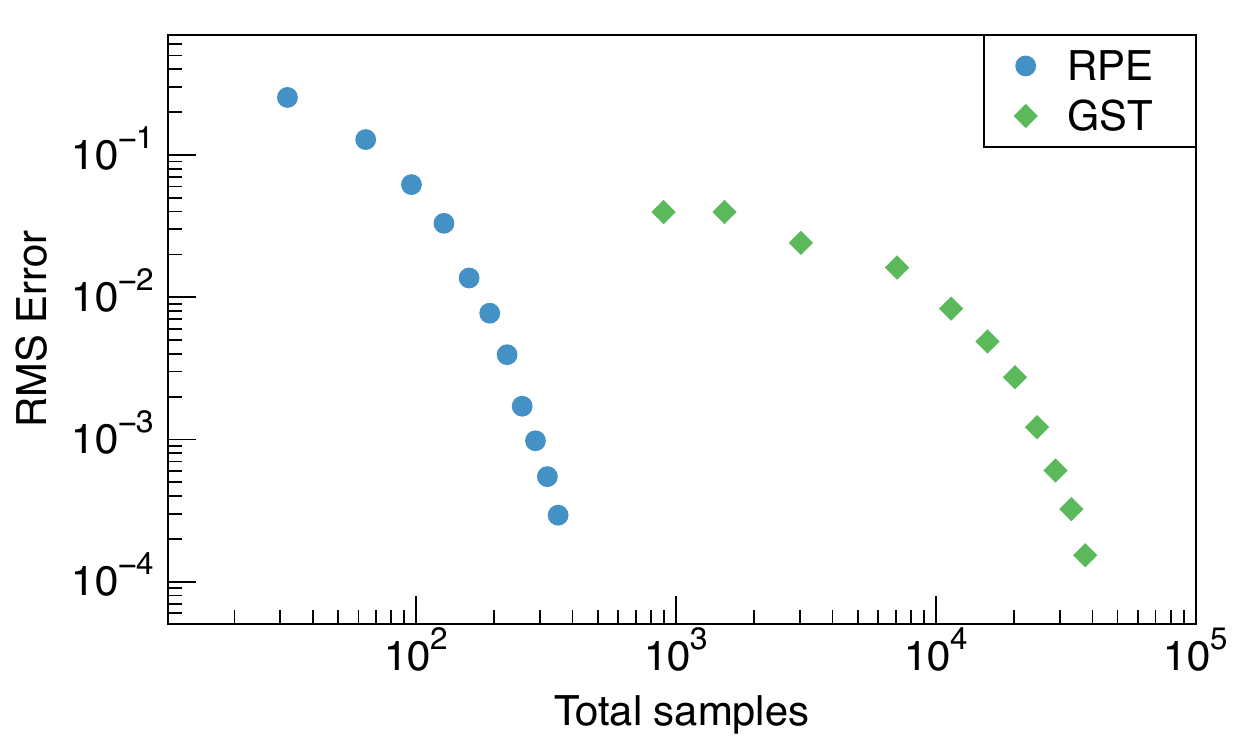}
\caption{\label{fig:Epsilon2}
(Color online) Root mean squared error vs total number of samples for RPE and GST estimates of the angle $\epsilon$ from the subsampled 
datasets with $N=16$.  This plot is analogous to the plot given in Fig. 5; for further details, see caption of that figure.}
\end{figure}

\bibliographystyle{apsrev4-1}
\bibliography{RPE}

\end{document}